\documentclass[12pt]{article}
\usepackage[T1]{fontenc}
\usepackage{amssymb,amsthm,amsmath,mathrsfs,bm}

\begin{document}
\title{ Two-Componet Coupled KdV Equations and its Connection with the Generalized Harry Dym Equations}
\author{ Ziemowit Popowicz}

\maketitle

\begin{center}{
Institute of Theoretical Physics, University of Wroc\l aw,

Wroc\l aw pl. M. Borna 9, 50-205 Wroc\l aw Poland, ziemek@ift.uni.wroc.pl}
\end{center}
\vspace{0.8cm} 

\begin{abstract}
It is shown that, three  different Lax operators in the Dym hierarchy, produce three generalized coupled 
Harry Dym equations. These equations transform, via the reciprocal link,  to the  coupled two-component KdV system. 
The first equation gives us known  integrable two-component KdV system while the second  reduces to the  known  symmetrical two-component KdV equation. The last one reduces to the Drienfeld-Sokolov equation. This approach gives us new Lax representation for these equations.

\end{abstract}

\newpage

\section{Introduction}
There are many different methods  of the classification of integrable equations. 
The most popular is the utilizations of the conservation laws and the generalized symmetry. These methods  
 led to the discovery of many new integrable systems \cite{Micha1,Micha2}, both S-integrable and C-integrable in Calogero's terminology \cite{Calo}. 

On the other side these methods have been used as well as to the classifications of the two-component coupled  KdV type equations. For example Foursov in 2003 \cite{fours1,fours2}   tested the integrability considering the following system of equations
\begin{equation}
 u_t = F(u,v), ~~~~~ v_t=G(u,v)
\end{equation}
where $F(u,v) = F(u,v,u_x,v_x,u_{xx}\dots)$ denotes  a differentail polynomial function of $u$ and $v$ .

As the result Foursov presented five non-symmetrical \cite{fours1} and 12  symmetrical \cite{fours2} two-component coupled KdV systems which   possesses several higher-order symmetries and conserved quantities. The symmetrical system  is such  where $G(u,v) = F(v,u)$. 

The first three non-symmetrical  systems in this classification are known to be integrable equations  
\begin{eqnarray}
 u_t &=&  u_{xxx} + 6uu_{x} -12 vv_x, ~~~~~ v_t = -2v_{xxx} - 6uv_x \\
 u_t &=&  u_{xxx} + 3uu_{x} + 3vv_x,   ~~~~~ v_t = u_xv + uv_x \\
 u_t &=&  u_{xxx} + 2vu_{x} + uv_x,  ~~~~~~~~v_t = uu_x
\end{eqnarray}

The first equation  is the Hirota - Satsuma system \cite{hirek}, second is the Ito system \cite{itek},  third is the rescaled Drinfeld - Sokolov equation \cite{sokol} . 

The fourth system 
\begin{eqnarray}\label{fours_1}
  u_t &=&  u_{xxx} + v_{xxx} + 2vu_{x} + 2uv_x  \\ \nonumber
 v_t &=& v_{xxx} - 9u u_x + 6 vu_x + 3uv_x + 2vv_x
\end{eqnarray}
possesses several  generalized symmetries   and several higher order   conserved densities  and  therefore  Foursov conjectured that it  is integrable and should possesses infinitely many generalized symmetries. 

The last system in this classification  
\begin{eqnarray}\label{fours1}
  u_t &=&  4u_{xxx} +3v_{xxx} + 4uu_x + vu_x + 2uv_x  \\ \nonumber
 v_t &=& 3u_{xxx} + v_{xxx} -4vu_x - 2uv_x - 2vv_x
\end{eqnarray}
  is integrable and possesses the Lax pair representations  
\begin{equation} \label{laxik}
 L=(\partial_{xxx} + \frac{2}{3}u\partial_x + \frac{1}{3}u_x)(\partial_{xx} - \frac{1}{3}v), ~~~~
\frac{\partial L}{\partial t} = 5 [L,L^{3/5}_{+}]
\end{equation}
and has been first considered many years ago by Drinfeld  and Sokolov \cite{sokol}  and rediscovered by S. Sakovich \cite{sak1}

In this paper we will discuss the problem how the coupled two-component KdV systems are connected 
with  the  generalized  two component  Harry Dym systems.
It is very well known that, Korteweg de Vries equation is connected with the Harry Dym equation \cite{ibra,her}.
Strictly speaking the Harry Dym equation is reciprocaly  linked   with  the Korteweg de Vries  equation. 
However the problem how  it is possible to transforms five  of the mentioned  two-component coupled  KdV systems to some new generalization of the  Harry Dym equations is an open problem. 
Such equivalence  exists only for the Hirota-Satsuma equation \cite{pop1,sak2}.

The one possible manner to find  such connections  is to  apply the inverse reciprocal link to the coupled KdV systems. However  in this procedure we have to assume the most general ansatz on the coupled modified KdV system. Next from these modified KdV functions we have to construct  new functions, assuming once more the most general ansats, in order to  yields  the generalized Harry Dym equation. This approach lead us to the algebraic system which is impossible to solve. 

The next manner is to first find some  generalization  of the coupled  Harry Dym equations and next construct the reciprocal link. If we progress  it we expect to obtain the coupled KdV type systems. At the moment we known two different two-component generalization of the Harry Dym equation. The first  considered in \cite{pop1,sak2} is connected with the Hirota-Satsuma equation. The second, defined  in \cite{fordy},  does not lead us to the coupled two component KdV system as we checked using the reciprocal link.

In this paper we use the second manner and present three Lax representations which produce new   generalizations of the  Harry Dym equation. Applying the reciprocal link to these equations we obtained the equations \ref{fours1}, the Drinfeld-Sokolv  and one known symmetrical coupled KdV equation listed in \cite{fours2}. 

The idea of the construction of the   Lax representation  for the generalized Harry Dym system follows from the observation that the Lax operator \ref{laxik}  which,  generates the system \ref{fours1},  is exactly the product of two operators. The first  is the Lax operator,  which produces the Kupershmidt equation,  while the second is the KdV Lax operator. It is known that analogon of the Lax operator for the Kupershmidt equation, in the Dym hierarchy,  is  third order operator while analogon of the Lax operator for  the KdV equation is the Harry Dym Lax operator. If we consider the product of these two operators we expect to obtain some new generalization of the Harry Dym equation and in the next,  by application of  the reciprocal link, new interacted two-component KdV equations. 

In principle we should consider three  different  operators because the Harry Dym equation possesses  two different Lax operators  the standard \cite{konop}  and recently discovered the nonstandard \cite{pop2}. We show that the Lax operator
constructed as the product of third order and standard Harry Dym operators leads us to the  equation \ref{fours1} . The operator constructed as the product  of third order operator and non-standard Harry operator gives us known symmetrical two-component KdV equation. The Lax operator constructed as the product of standard and nonstandard Lax operator of Harry Dym leads us to the Drinfeld-Sokolov equation. 

The paper is oragnized as follows. In the second  section  we present three Lax operators which generate 
three generalization of the Harry Dym equation. The third section describes the reciprocal link from generalized  Harry Dym equations to the coupled two-component KdV equations. 
In the fourth section we compare the Lax representation obtained using our approach with the known 
Lax representation for the coupled two-component KdV equations. 

integrability properties of new coupled KdV equations. The last section contains concluding remarks.

\section{New generalization of the Harry Dym equation} 

We meet three equivalent expression on the Harry Dym (HD) equation in the literature  \cite{her,brun}
\begin{equation}
w_t = (w^{-1/2})_{xxx},~~~~
u_t = ((u_{xx})^{-1/2})_x, ~~~~~~~
v_t = v^3 v_{xxx}
\end{equation}
where $v=-2^{1/3}w^{-1/2}$ and $u_{xx}=w$ respectively.

This equation have been discovered by H. Dym and M. Kruskal in 1975 \cite{krus} and share many 
of the properties typical of the soliton equations as for example it has  a Bi-Hamiltonian structure 
\begin{equation}
 w_t = {\cal D}_1 \frac{\delta H_1}{\delta w} = {\cal D}_2 \frac{\delta H_2}{\delta w} 
\end{equation}
where 
\begin{eqnarray}\label{biha}
 {\cal D}_1 &=& \partial^3, \quad \quad ~~~~~~~~~~~~~~~ {\cal D}_2 = \partial w + w \partial   \\ \nonumber
H_1 &=& 2\int ~dx ~w^{-1/2} , \quad \quad H_2=8\int ~dx ~ w^{-5/2} w_x^2
\end{eqnarray}

 and an infinite number of conservation laws and infinitely many symmetries \cite{brun}.
This equation is connected with the Korteweg de Vries (KdV) equation via the reciprocal transformation \cite{her,ibra}.

The HD equation follows from the following  Lax representation.
\begin{equation}
 L_{s}=\frac{1}{w} \partial^2, ~~~~~~~~~~
L_{s,t}=2[L_s,(L_s^{3/2})_{\geq 2}]
\end{equation}
where  subscript $\geq 2$ denotes the projection to the part with the powers greater or equal to $\partial ^2$.

On the other side there exists the second Lax operator for the Harry Dym equation, the nonstandard one, recently discovered in \cite{pop2} , 
\begin{equation}\label{nonlax}
 L_{ns}=w^{-1/4}\partial^{-1}w^{-1/4}\partial^2, ~~~~~~
L_{ns,t}=[L_{ns},(L_{ns}^{3})_{\geq 2}]/2
\end{equation}

Now taking into account that the system of equation \ref{fours1} has the Lax representation \ref{laxik} in which the Lax operator is factorized as a product of  the Lax operator of the Kupershmidt equation and the 
Lax operator of the Korteweg de Vries  equation  let us consider   four    Lax operators $L_s,L_{sn},L_{HS},L_{DS}$
\begin{eqnarray} 
 L &=& w^3\partial_{xxx} + k_0w^2w_x\partial_x   \\ \nonumber 
L_s &=& L v^2\partial^2, ~~~~~  L_{SD} = L v\partial^{-1}v\partial^2 \\ \nonumber 
L_{HS} &=& w\partial^2 u\partial^2 , ~~~~~
L_{DS}  = w^{-1/2}\partial^{-1}w^{-1/2} \partial^2 \frac{1}{u}\partial^2
\end{eqnarray}
The operator $L_{HS}$ has been considered in \cite{pop1} and leads us to the Hirota-Satsuma equation. From that reasosn we will not consider it in this paper. 

Let us mention that  the operator $L$ generates by 
\begin{equation}
 L_t = [L^{5/3}_{\geq 2},L]
\end{equation}
the fifth order equation which,  could be transformed by reciprocal link, to the Kupershmidt or  Sawada-Kotera equation 
 for $k_0=3$ or $k=\frac{3}{2}$ respectively.

The time evolution of the $\hat L_s, \hat L_{DS}$  and $L_{DS}$ 
\begin{equation} 
  L_{SD,t} = [( L_{SD}^{3/5})_{\geq 2},L_{SD}, ~~~~ L_{s,t} = [( L_{s}^{3/5})_{\geq 2}, L_{s}],~~~
L_{DS,t} =   [(L_{DS}^{3})_{\geq 2},L_{DS}] 
\end{equation}
produces the consistent solution only for $k_0=\frac{3}{2}$ 

For $L_{SD}$ operator we obtained 
\begin{equation}
 w_t = \frac{2}{3} w^{5/2} \Big ( w^{3/2}(v^{6/5}w^{-6/5})_x \Big )_{xx}, ~~ 
v_t = \frac{1}{4} v^3 (w^{9/5}v^{-4/5})_{xxx}
\end{equation}

while for $L_{s}$
\begin{equation}
w_t = \frac{2}{3}w^{5/2} \Big (w^{3/2}(v^{3/2}w^{-3/4})_x \Big )_{xx} , ~~~
v_t = v^3  ( w^{3/4}v_{xx} v^{-3/4})_{x} 
\end{equation}

For the $L_{DS}$ operator we obtained
\begin{equation}\label{laxds}
u_t = -\frac{1}{2}\Big (\frac{1}{w}\Big )_{xxx}, ~~~~w_t = -2 \Big ( \frac{\sqrt{w}}{u} \big (\frac{1}{\sqrt{w}}\big )_{xx} \Big )_x 
\end{equation}

\section{ReciprocaL link} 

Introducing the parametriaztion of the function $w,v$ for the $L_{SD}$ operator as 
\begin{equation}
w=ae^{b}, ~~~~~~~~~  v=ae^{-3b/2}
\end{equation}
and for the $L_{s}$ operator as
\begin{equation}
 w=ae^{b}, ~~~~~~~~~~v={\sqrt a}e^{-3b/2}
\end{equation}
and for the $L_{DS}$ operator as 
\begin{equation}\label{aba}
u=\frac{1}{a} e^b \quad \quad w=\frac{1}{a^2} e^{-b}
\end{equation}

we obtained that  the $L_{SD}$ operator generates 
\begin{eqnarray}
a_t &=& \frac{1}{10} (-a_{xxx} a^3 + 9 b_{xxx} a^3 + 27 b_{xx}( a_x a^{3} -3b_{x}a^3)  -\\ \nonumber 
&& \hspace{2cm}  81b_x^{2}a_xa^{3} -  9b_x( a_{xx} a^3 - a_x^2 a^2)), \\ \nonumber
b_t&=& \frac{1}{10}(a_{xxx}a^2 +11b_{xxx}a^3 +b_{xx}( 33a_{x}a^{2} - 9b_xa^{3}) + \\ \nonumber 
&&\hspace{1cm}  45b_{x}^{3}a^3 - 9b_{x}^{2} a_{x}a^{2} + b_x(21a_{xx}a^{2}+ 6a_{x}^2a))
\end{eqnarray}
while the  $L_{s}$ operator gives us 
\begin{eqnarray}
a_t &=&\frac{1}{4} ( a_{xxx}a^3 - 9b_{xxx}a^4 - 27b_{xx}(a_xa^3 -b_xa^4) +27b_{x}^2a_xa^3 - \\ \nonumber  
&& \hspace{2cm}  9b_x(a_{xx}a^3 + a_x^2a^2)) \\ \nonumber 
b_t &=&  \frac{1}{4}(-a_{xxx}a^2 + b_{xxx}a^3 + 3b_{xx}(a_xa^2 + 3b_xa^3) -18b_x^3 a^3 + \\ \nonumber
&& \hspace{1cm} 9b_x^2a_xa^2 - 3b_x(a_{xx}a^2 - a_x^2a))
\end{eqnarray}
and for the $L_{DS}$ operator 
\begin{eqnarray}
 a_t &=& \frac{a^2}{2} \big (4a_{xx}a + b_x^2 a^2 + 2(b_xa^2)_x \big )_x \\ \nonumber 
b_t &=& \frac{1}{2}\big (2a_x^3 - b_x^3a^3 - 2b_xa^2(b_xa_x + a_{xx}) - (2a_x^2a+b_x^2a^3+2b_xa_xa^2)_x\big )
\end{eqnarray}

We use  the reciprocal transformation where now $x,a$ and $b$ are defined as 
\begin{equation}\label{reci1}
 x=p(y,t), \quad a(x,t)=p(y,t)_y, \quad b(x,t) = q(y,t)
\end{equation}
in order to find the time evolution of $p,q$.

For $L_{SD}$ case we obtained 
\begin{eqnarray}\label{pq}
 p_t &=& \frac{1}{20}( -2 p_{yyy} + 3\frac{p_{yy}^2}{p_y} + p_y(18q_{yy} - 81 q_y^2) )\\ \nonumber 
q_t &=&\frac{1}{20}\Big ( 2\frac{p_{4y}}{p_{y}}   - p_{yy}(8\frac{p_{yyy}}{p_y^{2}} - 6\frac{p_{yy}^{2}}{p_{y}})   +22q_{yyy} + q_y(9q_y +18\frac{p_{yyy}}{p_{y}} - 27\frac{p_{yy}}{p_y^2}) \Big) 
\end{eqnarray}
while for $L_{s}$ 
\begin{eqnarray}
p_t &=& \frac{1}{8}(2p_{yyy} - 3\frac{p_{yy}^2}{p_y} -p_y(18q_{yy}-27q_y^2)) \\ \nonumber 
q_t &=& \frac{1}{8}(-2\frac{p_{4y}}{p_y} + p_{yy}(8\frac{p_{yyy}}{p_y^2} - 6\frac{p_{yy}}{p_y^3}) + 2q_{yyy} -
 q_y(9q_y + 6\frac{p_{yyy}}{p_y} - 9\frac{p_{yy}^2}{p_y^2}))
\end{eqnarray}
and for the $L_{DS}$ operator 
\begin{eqnarray}\label{pq}
 p_t &=& \frac{1}{4}( 4 p_{yyy} - 4\frac{p_{yy}^2}{p_y} + 2q_{yy}p_{y} + q_y^2p_y + 2q_yp_{yy} )\\ \nonumber 
q_t &=&\frac{1}{4}\Big ( - 8p_{yyy}p_{yy}p_{y}^{-2}  + 8p_{yy}^3p_y^{-3} - 2q_{yy}q_y - q_y^3 - 
4p_y^{-1}(q_{yy}p_{yy} + q_y^2p_{yy} + q_yp_{yyy}) \Big) 
\end{eqnarray}

To verify this one can use the identities
\begin{equation}\label{reci2}
 \partial_x = \frac{1}{a} \partial_y, \quad a_t = p_{y,t} - \frac{ p_{yy}p_{t} }{ p_{y}}, \quad  b_t= q_t -  \frac{q_{y}p_{t}}{ p_{y}}
\end{equation}
Next we apply  the transformation 
\begin{equation}\label{fg}
 f=\frac{p_{yy}}{p_y} , \quad g=q_y
\end{equation}
from which we conclude that for the $L_{SD}$ operator we obtained 
\begin{eqnarray}\label{mkdv0}
 f_t &=& \frac{1}{20} \Big (-2f_{yy} + f^3 + 18g_{yy} + 18g_yf-162g g_y - 81g^2f \Big )_y \\ \nonumber 
  g_t &=&  \frac{1}{20} \Big ( 2f_{yy} - 2f_yf + 22g_{yy} + 9g^3 + 18gf_y - 9gf^2  \Big )_y
\end{eqnarray}
while for the $L_{s}$ operator 

\begin{eqnarray}
f_t &=&\frac{1}{8}(2f_{yy} - f^3 - 18g^2 - 18g_yf + 54g_yg + 27g^2f)_y \\ \nonumber
g_t &=& \frac{1}{8}(-2f_{yy} + f_yf + 2g_{yy} - 9g^3 - 6gf_y + 3gf^2)_y
\end{eqnarray}

and for the $L_{DS}$ operator 
\begin{eqnarray}\label{mkdv}
 f_t &=& \frac{1}{4} \Big ( ( 4f_{y} + 2g_{y} +2f^2 + g^2 + 2gf)_y + 2g_yf + g^2f + 2gf^2\Big )_y \\ \nonumber 
 g_t &=& - \frac{1}{4} \Big ( (4f^2 + g^2 + 4gf_y)_y  + g^3 + 4g^2f + 4gf^2 \Big )_y
\end{eqnarray}

If we apply the Miura-type transformation for the $L_{SD}$ case 
\begin{eqnarray}\label{mirek0}
 u &=& \frac{1}{8} (8 f_y +3f^2 + 27g^2 - 6fg) \\ \nonumber 
 v &=& \frac{1}{24} (24 g_y - f^2 - 9g^2 + 18 fg)
\end{eqnarray}
then the system of equation \ref{mkdv} reduces to 
\begin{eqnarray}\label{kodek}
 u_t &=& \frac{1}{20} (- 2u_{yyy} + 18v_{yyy} + 9uu_y - 9u v_y - 405v v_y-27v u_y) \\ \nonumber
 v_t &=& \frac{1}{20} (2u_{yyy} + 22v_{yyy} -5u u_y -27u v_y + 9v v_y - 9v u_y)
\end{eqnarray}
If we apply the linear transformation of the variables $u,v$ and apply the scale of the time  
\begin{equation} 
 t \Rightarrow \frac{t}{4}, ~~~ u \Rightarrow \frac{v-u}{3}, ~~~~~ v \Rightarrow -\frac{u+v}{9}
\end{equation} 
then the system \ref{kodek} reduces to the Drinfeld-Sokolov equation. 

Applying different Miura-type transformation for the $L_{s}$ case 
\begin{equation}\label{mirek1}
 u = \frac{1}{2} (2 f_y - f^2 ), ~~~~~~~~ v = \frac{1}{2} (2 g_y - 3g^2 )
\end{equation}
we obtained 
\begin{eqnarray}\label{symkdv}
 u_t &=& \frac{1}{4}(u_{yyy} +3u u_y - 9v_{yyy} - 18u v_y - 9v u_y) \\ \nonumber
 v_t &=& \frac{1}{4}(-u_{yyy}+ v_{yyy} - 3uv_y -6v u_y + 9vv_y )
\end{eqnarray}

Applying the scaling $u \Rightarrow \frac{2}{3} u, v \Rightarrow \frac{2}{9} v$ the system of equation \ref{symkdv} reduces to the symmetrical form 
\begin{equation} 
 u_t =-2(-\frac{1}{2} u_{yyy} - u u_y + \frac{3}{2} v_{yyy} + 2v_y u + vu_y).
\end{equation}
It is one of the rescaled symmetrical  coupled two-component  KdV system considered by Foursov  \cite{fours2} \footnote{the equation 4.9 in \cite{fours1} where $\alpha = 1$.} .

If we apply the Miura-type transformation  
\begin{equation}\label{mirek}
 s= f_y -\frac{1}{2}f^2,\quad z=g_y +f^2+\frac{1}{2}g^2 + fg
\end{equation}
for the $L_{DS}$ case, then the system of equation \ref{mkdv} reduces to 
\begin{eqnarray}
 s_t &=& \frac{1}{2} \Big ( ( 2s_{yy} +z_{yy} +3s^2 + sz )_x +z_xs \Big )_y  \\ 
 z_t &=&-\frac{3}{4} (4s^2 + 2zs + z^2)_y.
\end{eqnarray}
If we further shift the function $ s \Rightarrow s - z/2$ we obtain 
\begin{eqnarray}
 s_t &=& \frac{1}{2}  (2s_{xxx} - z_x s - 2z s_x  )  \\ 
 z_t &=&-6s_ys.
\end{eqnarray}
 It is exactly the first Drinfeld - Sokolov equation DS1  if we apply   the scaling $s \Rightarrow s/2,
z \Rightarrow -z, \partial_t \Rightarrow \partial_t/2 $.

\section{The Lax representations of the coupled two-component  KdV type systems} 

The known Lax representation of the Drienfeld-Sokolov \cite{sokol} is 
\begin{equation} 
 L=(\partial^3 +(s+z)\partial + \frac{1}{2}(s+z)_x)(\partial^3 + (s-z)\partial + \frac{1}{2}(s-z)_x), ~~~~ L_t=[L^{1/2}_{\geq 0},L]
\end{equation}
while for the system \ref{fours1} is \ref{laxik}. 

We show that our approach produces quite different Lax represenation for these equations.

First let us consider the $L_{DS}$ operator for which we apply the gauge transformation 

\begin{equation}
\hat L_{DS} = e^{-b/2} L_{DS} e^{b/2} 
\end{equation}
and next the reciprocal transformation and  once more the gauge transformation 
\begin{equation}
 \hat L_{DS}  \Rightarrow \frac{1}{p_y} \hat L_{DS} p_y
\end{equation}
Finally if we apply  the   Miura-type transformation \ref{mirek}  we obtain 
\begin{eqnarray}
 \hat L_{DS}  &=& \partial^3 + (3s+2z)\partial + \partial^{-1}\big (s_{yy} + \frac{1}{2}z_{yy} - \frac{1}{4}z^2 - \frac{1}{4} zs \big ) - \\ \nonumber 
&& \frac{3}{2} (f+g)^2 \partial + \partial^{-1} \Big ( (f+g) \Big ( \frac{1}{2} (s+\frac{1}{2}z)(f+g) - s_y - 
\frac{1}{2} z_y\Big ) \Big )
\end{eqnarray}
Strictly speaking we obtained the Lax operator which generates the time evolution of the function $f,g$ and implicitly the Drinfeld-Sokolv  equation. Interestingly this operator produces also the conserved quantities for this equation. It can be easily seen for the first quantities 
\begin{equation}
 H_0 = Res(L) = \int ~dy~ s^2 + \frac{1}{4}z^2 + zs.
\end{equation}

For the $L_{s}$ operator making the similar transformations we obtined 
\begin{eqnarray} \nonumber 
 \hat L_s &=& \partial_{4x} + (u+3v)\partial_{xx} + \frac{3}{2}(u+3v)_x\partial + \frac{1}{2}(u+3v)_{xx} + 
\\ \nonumber
&& \frac{1}{4}(u-3v)^2 + \frac{1}{4}(u-3v)_x\partial^{-1}(u-3v)
\end{eqnarray}

For the $L_{DS}$ operator making the similar transformations we obtined 

\begin{eqnarray}  \nonumber 
 \hat L_{DS} &=& e^{-\int dy((f + 3g)/2)} L_{DS} e^{\int dy((f+3g)/2)} = 
 (\partial_{yyy} - 3\partial_{yy} g + \frac{1}{4}\partial_y(2 f_y - f^2 + 6g_y + 9g^2)) \\ 
&& \hspace{2cm} ( \partial_{yy} + 3g\partial_y + \frac{1}{4}(2f_y - f^2 + 6g_y + 9g^2)).
\end{eqnarray}

The time evolution of $\hat L_{DS}$
\begin{equation}
 \hat L_{DS,t}  = [(\hat L_{DS}^{3/5})_{+},\hat L_{DS}]
\end{equation}
generates the  coupled mKdV equation \ref{mkdv0} and in the implicit form also the system of \ref{fours1}.  However  $\hat L_{s}$   can not be rewritten, after 
applications of the Miura transformation \ref{mirek0}, purely in terms of the local KdV variables $u,v$ and its derivatives.
As we checked the residuum formula of the $\hat L$ generates the conserved quantity  for the MKdV equation as well as for  the coupled KdV system \ref{kodek}. Indeed the first two nontrivial conserved laws obtained from $\hat L$ operator are
\begin{eqnarray} 
 G_2 &=& res(\hat L^{1/5}) = \int  (f^2 + 9g^2)dy \\ \nonumber 
 G_4 &=& res( \hat L^{3/5}) = \int  (4f_{yy}f - f^4 - 72g^2f -396g_{yy}g -36g_yf^2 +\\ \nonumber 
&& \hspace{3cm}  648g_ygf - 81g^4 + 162g^2f^2)dy
\end{eqnarray}
Here the lower  index in $G,H$ denotes the KdV weight of the function e.g $[u]=[v]=2,[f]=[g]=1,[\partial]=1$.
In order to obtain the conserved quatities for the equation \ref{fours1} from the $\hat L_s$ operator 
let us apply the Miura transformation \ref{mirek} and rewrite these quantities as 
\begin{eqnarray} 
 G_2 &=& \int (f^2+9g^2 + 3f_y + 3g_y) dy = 3\int (u+v) dy = H_2\\ \nonumber
 G_4 &=& \int dy (u^2 - 99v^2 - 18uv).
\end{eqnarray}
In the similar way it is possible to obtain the  higher order conserved densities.

\section{Concluding remarks}
In this paper we presented three  different generalizations of the Harry Dym  equation. These equations have been obtained 
using  different Lax operators in the Harry Dym hierarchy. Using the reciprocal link we showed that our equations are reduced to the coupled two-component KdV equations. However how remaining  coupled two-component KdV equations listed in \cite{fours1,fours2} are connected with the generalized Harry Dym equations is still an open problem.


\begin{thebibliography}{99}

\bibitem{Micha1} A. Mikhailov, A. Shabat, V. Sokolov \textit{The symmetry approach to classificatiobn of integrable equations} in \textit{What is Integrability} edited by. V. Zakharov (Springer-Verlag 1991) 115-184.

\bibitem{Micha2} A. Mikhailov, A. Shabat, R. Yamilov \textit{The symmetry  approach to the classification of nonlinear equations. Complete lists of integrable systems} Russ. Math. Surveys \textbf{42} (1987) 115-184.

\bibitem{Calo} F. Calogero \textit{Why are certain nonlinear PDEs both widely applicabe and integrable}
in \textit{What is Integrability} edited by. V. Zakharov (Springer-Verlag 1991) 1-62.


\bibitem{fours1} M. Foursov \textit{Towards the complete classification of homogenous two-component integrable equations}
J. Math. Phys. 44 (2003) 3088-3096. 


\bibitem{fours2} M. Foursov \textit{On integrable coupled KdV-type systems}
Inverse Problems \textbf{16} (2000), 259 - 274.

\bibitem{hirek} R. Hirota, J. Satsuma  \textit{Soliton solutions of a coupled Korteweg-de Vries equation} 
Phys. Lett 85A 407-408.

\bibitem{itek}  M. Ito, \textit{Symmetries and conservation laws of a coupled nonlinear wave equation}, Phys. Lett. A, 1982, V.91, 335-338.

\bibitem{sokol} V.  Drinfeld, Sokolov V, \textit{New evolutionary equations possessing an (L,A)-pair}, Trudy Sem. S.L. Soboleva (1981), no.2 5-9 (in Russian).

\bibitem{sak1} S.Sakovich  \textit{Coupled KdV Equations of Hirota-Satsuma Type}
Journal of Nonlinear Mathematical Physics 1999, V.6 N 3  255 -262, ibid 
Addendum to: \textit{Coupled KdV Equations of Hirota-Satsuma Type} 
Journal of Nonlinear Mathematical Physics Volume 6, Number 2 (2001), 311-312.




\bibitem{ibra} N.H Ibragimov \textit{Sur l'\'{e}quivalence des \'{e}quations d'\'{e}volution, qui 
admettent une alg\'{e}bre de Lie-B\"{a}cklund infinie}   C. R. Acad. Sci., Paris (1981) 293 657-60.

\bibitem{her} W. Hereman, P. P. Banerjee and M. R. Chatterjee, 
\textit{On the Nonlocal Equations and Nonlocal Charges Associated with the Harry Dym  Hierarchy
Korteweg-de Vries equation} J. Phys. A 22 (1989) 241.

\bibitem{pop1}  Z. Popowicz \textit{The Generalized Harry Dym Equation} Phys. Lett. \textbf{A} 317  260-264 (2003).


\bibitem{sak2} S. Sakovich  \textit{Transformation of a generalized HD equation into the Hirota-Satsuma system} Phys. Lett.  A321 (2004) 253-254.

\bibitem{fordy} M. Antonowicz, A. Fordy \textit{Coupled Harry Dym equations with multi-Hamiltonian structures} 
J. Phys. A: Math. Gen. \textbf{21} (1988) L269-L275.

\bibitem{konop} B. Konopelchenko, W.Oevel; \textit{An r-Matrix  Approach to Nonstandard Classes of Integrable Equations}  Publ. Rims Kyoto Unive. 29 (1993) 58. 

\bibitem{pop2} K. Tian, Z. Popowicz, Q. Liu  \textit{A non-standard Lax formulation of the Harry Dym hierarchy and its supersymmetric extension}  
 J. Phys A:Math.Theor. 45 (2012) 122001 (8pp).

\bibitem{krus} M.D. Kruskal Lecture Notes in Physics vol.38, Springer Berlin 1975, p. 310.

\bibitem{brun} 
 J.C. Brunelli, G.A.T.F. da Costa
\textit{ On the Nonlocal Equations and Nonlocal Charges Associated with the Harry Dym Hierarchy}
J.Math. Phys. 43 (2002) 6116-6128.


\end{thebibliography}
\end{document}